\documentclass[10pt,letterpaper]{article}
\usepackage{opex3}
\usepackage{graphicx}
\usepackage{cite}
\usepackage{color}
\usepackage{amsmath}
\usepackage{here}
\usepackage{epstopdf}

\newcommand{\be}{\begin{equation}}
\newcommand{\ee}{\end{equation}}
\newcommand{\bea}{\begin{eqnarray}}
\newcommand{\eea}{\end{eqnarray}}

\newcommand\rpict[1]{\ref{#1}}

\begin{document}

\begin{sloppy}

\title{Efficient excitation and tuning of toroidal dipoles within individual homogenous nanoparticles}

\author{Wei Liu,$^{1,*}$ Jianhua Shi,$^{1}$ Bing Lei,$^{1}$ Haojun Hu,$^{1}$ and Andrey E. Miroshnichenko$^2$}
\address{
$^1$College of Optoelectronic Science and Engineering, National University of Defense
Technology, Changsha, Hunan 410073, P. R. China.\\
$^2$Nonlinear Physics Centre, Centre for Ultrahigh-bandwidth Devices for Optical Systems (CUDOS), Research School of Physics and Engineering,  Australian National
University, Acton, ACT 0200, Australia.\\
$^*$Corresponding author: wei.liu.pku@gmail.com
}
%

\begin{abstract}
We revisit the fundamental topic of light scattering by single homogenous nanoparticles from the new perspective of excitation and manipulation of toroidal dipoles. It is revealed that besides within all-dielectric particles, toroidal dipoles can also be efficiently excited within homogenous metallic nanoparticles. Moreover, we show that those toroidal dipoles excited can be spectrally tuned through adjusting the radial anisotropy parameters of the materials, which paves the way for further more flexible manipulations of the toroidal responses within photonic systems. The study into toroidal multipole excitation and tuning within nanoparticles deepens our understanding of the seminal problem of light scattering, and may incubate many scattering related fundamental researches and applications.
\end{abstract}



\section{Introduction}

Stimulated by the recent experimental demonstration of toroidal dipoles (TDs) excited on the platform of metamaterials consisting of specially organized split ring resonators~\cite{Kaelberer2010_Science}, studies into TDs and more generally toroidal multipoles (TMs) within photonic nano-systems have attached surging attention~\cite{dong2012toroidal,Huang2012_OE,Ogut2012_NL,Fedotov2013_SR,Fan2013_lowloss,savinov2014toroidal,Xiao2014_AP_core,Basharin2014_arXiv,miroshnichenko2014seeing,Liu2014_arxiv,Bakker2015_NL_Magnetic,Liu2015_OL_invisible}. Compared to conventional  electric and magnetic multipoles, TMs can be viewed as poloidal currents flowing on the surface of a torus along its meridians, or as a series of magnetic dipoles (enclosed circulating currents) aligned along enclosed paths~\cite{Dubovik1990_PR,Radescu2002_PRE}. For the near field in terms of charge-current distributions, TMs are contrastingly different from their electric and magnetic counterparts~\cite{Fedotov2013_SR,savinov2014toroidal}. In the far field however, TMs and conventional  electric multipoles are indistinguishable~\cite{Fedotov2013_SR,savinov2014toroidal}, which results in the situation that in various scattering theories concerning particle scattering, TMs are not considered as separate entities~\cite{jackson1962classical,Bohren1983_book}.  Generally under most circumstances, the contributions of TMs have been incorrectly attributed to conventional  electric multipoles, rendering TMs unfortunately as missing or hidden terms.

Following the first experimental demonstration of TDs~\cite{Kaelberer2010_Science}, TMs have been observed in various configurations, most of which (influenced by the configuration designed in Ref.~\cite{Kaelberer2010_Science}) are kind of toroid-shaped composite structures~\cite{dong2012toroidal,Huang2012_OE,Ogut2012_NL,Fedotov2013_SR,Fan2013_lowloss,Basharin2014_arXiv}.  This is understandable: the corresponding charge-current distribution of TMs is in a toroidal shape and naturally in such composite structures TMs can be efficiently excited. But this unfortunately leads to more or less a dogma that TMs can only be observed in carefully engineered compound structures and consequently there have been very few searches for the existence of TMs in simple individual structures~\cite{Xiao2014_AP_core}. Only recently it is shown decisively that TMs can be excited within single scattering particles of fundamental (spherical or cylindrical) shapes, which also have been proven to play indispensable roles for some exotic scattering properties~\cite{miroshnichenko2014seeing,Liu2014_arxiv,Liu2015_OL_invisible}. The incorporation of TMs extends significantly our understanding of the fundamental problem of light scattering by spherical and cylindrical particles. Moreover, considering that the scattering of spherical and cylindrical particles is the cornerstone of many scattering related fundamental studies and applications, the incorporation of TMs into scattering photonic systems can possibly spur further breakthroughs in sensing, nanoantennas, nanoscale lasing and photovoltaic devices~\cite{Novotny2012_book}.

Here in this article we restudy the seminal problem of plane wave light scattering by homogenous spherical nanoparticles, but from a new angle of excitation and manipulations of TDs.  It is demonstrated that, similar to all-dielectric particles, even a homogenous metallic spherical nanoparticle can surprisingly support TDs in the spectral region beyond the quasi-static regime. We further show that radial anisotropy can be employed to tune spectrally the TDs excited, providing extra freedom for TMs manipulations, which can possibly find significant applications in TMs related topics, especially those where near-field distributions of fields and currents play a crucial role.

\section{Methods and Expressions}

\subsection{Conventional  Multipole Expansions and Emerging Toroidal Multipoles}

To obtain the radiated fields of an arbitrarily distributed charge-current source, the general approach is to firstly calculate the retarded vector potentials and then get the associated electromagnetic fields from those potentials~\cite{jackson1962classical}. When the retardation effect within the charge-current source can be neglected (in other words the geometrical dimension of the source is far smaller than the effective wavelength and thus the quasi-static approximation can be applied), the vector potentials can be Taylor-expanded into multiple terms of different orders, from which the conventional  electric and magnetic multipoles can be deduced~\cite{jackson1962classical}. For example, the zeroth order expansion term corresponds to the conventional  electric dipole $\textbf{P}$ [the $\rm exp(-i\omega t+i\textbf{k}\cdot\textbf{r})$ notation has been adopted where $\textbf{k}$ and $\omega$ denote wave-vector and angular frequency respectively]:
\begin{equation}
\label{ED}
\textbf{P}= {1 \over { - i\omega }}\int d^3r\textbf{J}(\textbf{r}),
\end{equation}
and the first order expansion term includes both conventional  electric quadrupole and magnetic dipole $\textbf{M}$:
\begin{equation}
\label{MD}
\textbf{M} = {1 \over {2c}}\int {d^3 r\left[ {\textbf{r }\times \textbf{J}(\textbf{r})} \right]},
\end{equation}
where $c$ is the speed of light.  Since $\textbf{M}$ corresponds to the higher order term of the Taylor expansion, it results in more or less the mind-sets that: (1) \textbf{P} should be stronger than the \textbf{M} and (2) like most higher order modes~\cite{Liu2014_arXiv_Geometric}, \textbf{M} should be supported at higher frequencies than \textbf{P}.  Nevertheless, we should keep in mind that the Taylor-expansion of the vector potentials~\cite{jackson1962classical} is based on the quasi-static approximation, and thus when the precondition of this approximation does not exist, the above conclusion is not valid anymore. An outstanding example of this is the recent demonstration of optically-induced magnetic response of higher permittivity dielectric particles~\cite{Kuznetsov2012_SciRep,Evlyukhin2012_NL,Liu2014_CPB}, where $\textbf{M}$ is supported at lower frequencies and can be stronger than the electric dipole.

 The other consequence of the breakdown of the quasi-static approximation is that the conventional  electric and magnetic dipoles [Eqs.~(\ref{ED})-(\ref{MD})] and higher order multipoles do not form a complete set and thus other multipoles have to be included as correcting terms~\cite{Dubovik1990_PR,Radescu2002_PRE,savinov2014toroidal}. For dipolar responses, besides $\textbf{M}$ and $\textbf{P}$, the lowest order correcting term would be the toroidal dipole:
\begin{equation}
\label{TD}
\textbf{T}= {1 \over {10c}}\int d^3r[(\textbf{r} \cdot \textbf{J}(\textbf{r}))\textbf{r} - 2r^2 \textbf{J}].
\end{equation}
The scattered power of $\textbf{P}$ and $\textbf{T}$ are respectively:
\begin{equation}
\label{ED_TD_power}
\rm W_{\rm \textbf{P}}  = {{\mu _0 \omega ^4 } \over {12\pi c}}\left| \textbf{P} \right|^2, ~~\rm W_{\rm \textbf{T}}  = {{\mu _0 \omega ^4 k^2 } \over {12\pi c}}\left| \textbf{T}\right|^2,
\end{equation}
where $\mu _0$ is the vacuum permeability.

\begin{figure*}
\centerline{\includegraphics[width=13.5cm]{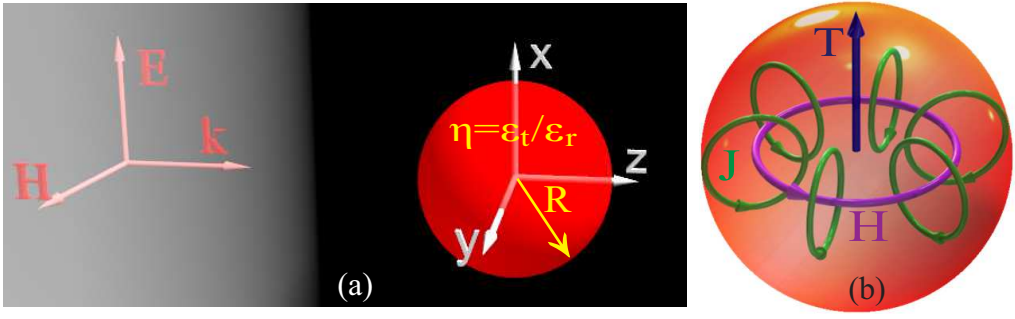}}\caption{(\small (a) Schematic illustration of the scattering of an incident plane wave by a spherical particle of radius R. The spherical particle can be homogenous and isotropic with reflective index $n$, or can be radially anisotropic with radial permittivity $\varepsilon_r$ and transverse permittivity $\varepsilon_t=n^2$. The anisotropy parameter is defined as $\eta=\varepsilon_t/\varepsilon_r$ ($\eta=1$ for isotropic spheres). The plane wave is polarized (in terms of electric field) along $x$ direction and is propagating along $z$ direction. (b)  Schematic illustration of the toroidal dipole excitation within the spherical particle. Both the current $\textbf{J}$ and magnetic field $\textbf{H}$ distributions have been shown.}
\label{fig1}
\end{figure*}

\subsection{Electric and magnetic multipoles associated with spherical harmonics of Mie scattering particles}
For spherical (both homogenous and multi-layered) particles with incident plane waves of electric field $\textbf{E}_0$, the scattered electric field can be expressed as~\cite{Bohren1983_book}£º
\begin{equation}
\label{harmonics}
\textbf{E}_{\rm s}(\textbf{r})  = \sum\limits_{m = 1}^\infty  {E_m } [ia_m \textbf{N}_{\rm m}(\textbf{r}) - b_m \textbf{M}_{\rm m}(\textbf{r})],
\end{equation}
where $a_m$ and $b_m$ are the Mie scattering coefficients; $\textbf{M}_{\rm m}$ and $\textbf{N}_{\rm m}$ are vector spherical harmonics; and $E_m  = i^m E_0 {{2m + 1} \over {m(m + 1)}}$. In the far field the scattered fields [Eq.(\ref{harmonics}) for $\textbf{r}\rightarrow\infty$] associated with $a_m$ [${E_m }ia_m \textbf{N}_{\rm m}(\textbf{r})$] and $b_m$ [${E_m }b_m \textbf{M}_{\rm m}(\textbf{r})$] are exactly the same as the far fields of conventional  electric and magnetic multipoles of order $m$, with conventional electric and magnetic dipoles $\textbf{P}$ and $\textbf{M}$ ($m=1$) defined in Eqs.(\ref{ED})-(\ref{MD}). For this far-field equivalence, it is natural to define another spherical harmonics based electric dipole as~\cite{Bohren1983_book,Doyle1989_optical,miroshnichenko2014seeing,Liu2014_arxiv,Wheeler2006_PRB}:
\begin{equation}
\label{conventioanl_dipole}
\textbf{P}(a_1)= \varepsilon _0 {{6\pi ia_1 } \over {k^3 }}\textbf{E}_0,
\end{equation}
where $\varepsilon _0$ is the vacuum permittivity, and the associated scattering power of $\textbf{P}(a_1)$ is~\cite{Radescu2002_PRE,miroshnichenko2014seeing,Liu2014_arxiv}:
\begin{equation}
\label{w_a1}
{\rm W_{\rm \textbf{P}(a_1)}} ={{3\pi \left| {E_0 } \right|^2 } \over {k\omega \mu _0 }},
\end{equation}
We note here that $\textbf{P}(a_1)$ has been obtained based on the far scattered field expansion and thus has included all the dipolar transverse magnetic components~\cite{Bohren1983_book,Liu2014_arxiv}. In contrast, the conventional  electric dipole $\textbf{P}$ shown in Eq.(\ref{ED})  has been obtained based on the near-field current integration, and here the current $\textbf{J}(\textbf{r})$ can be expressed as:

\begin{equation}
\label{current}
\textbf{J}(\textbf{r}) = -i\omega \varepsilon _0 [\varepsilon(\textbf{r})-1]\textbf{E}(\textbf{r}),
\end{equation}
where $\varepsilon(\textbf{r})$ is a scalar if the particle is isotropic and a tensor if the particle is radially anisotropic.  The conventional  electric dipole $\textbf{P}$ is only the lowest order expansion term of $\textbf{P}(a_1)$ shown in Eq.(\ref{conventioanl_dipole}). Similarly, the toroidal dipole  $\textbf{T}$ shown in Eq.(\ref{TD}) is the next higher order expansion of $\textbf{P}(a_1)$. So basically $\textbf{P}(a_1)$  can be expressed as~\cite{Dubovik1990_PR,Radescu2002_PRE,miroshnichenko2014seeing,Liu2014_arxiv}:
\begin{equation}
\label{dipole_expansion}
\textbf{P}(a_1)= \textbf{P} + ik\textbf{T}+\aleph,
\end{equation}
where $\aleph$ denote higher order correcting terms of $\textbf{P}(a_1)$ with respect to $kR$~\cite{Dubovik1990_PR,Radescu2002_PRE,Corbaton2015_arXiv_On,Corbaton2015_arXiv_Exact}. As is shown in Eq.(\ref{dipole_expansion}) the multipole expansion of the far scattered fields  based on spherical harmonics is complete [Eqs.(\ref{harmonics})-(\ref{conventioanl_dipole})] while the conventional  multipole expansion based on the integration of the near-field current is incomplete [Eqs.(\ref{ED})-(\ref{TD})]. As the normalized particle radius ($\alpha=kR$) is getting larger and larger (that is to say the particle radius R is getting larger and larger or the functioning wavelength is getting smaller or smaller), higher order correcting terms including the toroidal dipole $\textbf{T}$ would arise. That is to say, multipole expansions even including toroidal multipoles are still incomplete. For sufficiently larger $\alpha$, other higher order correcting terms must also be incorporated~\cite{Dubovik1990_PR,Radescu2002_PRE,miroshnichenko2014seeing,Liu2014_arxiv,Liu2015_OL_invisible,Corbaton2015_arXiv_On,Corbaton2015_arXiv_Exact}.

\section{Results and Discussions}

In Fig.~\ref{fig1}(a) the schematic of the scattering configuration is shown: the pane wave is scattered by a spherical particle, which could be homogenous and isotropic (metal or dielectric) with refractive index $n$ or radially anisotropic with radial permittivity $\varepsilon_r$ and transverse (perpendicular to the radial direction) permittivity $\varepsilon_t=n^2$. The anisotropy parameter is defined as $\eta=\varepsilon_t/\varepsilon_r$. The plane wave is polarized along $x$ direction and propagating along $z$ direction. In Fig.~\rpict{fig1}(b) we show the excited toroidal dipole $\textbf{T}$ within the particle by the magnetic field \textbf{H} and current distribution \textbf{J}.

\subsection{TD excitation within homogenous and isotropic dielectric particles}

\begin{figure*}
\centerline{\includegraphics[width=13.5cm]{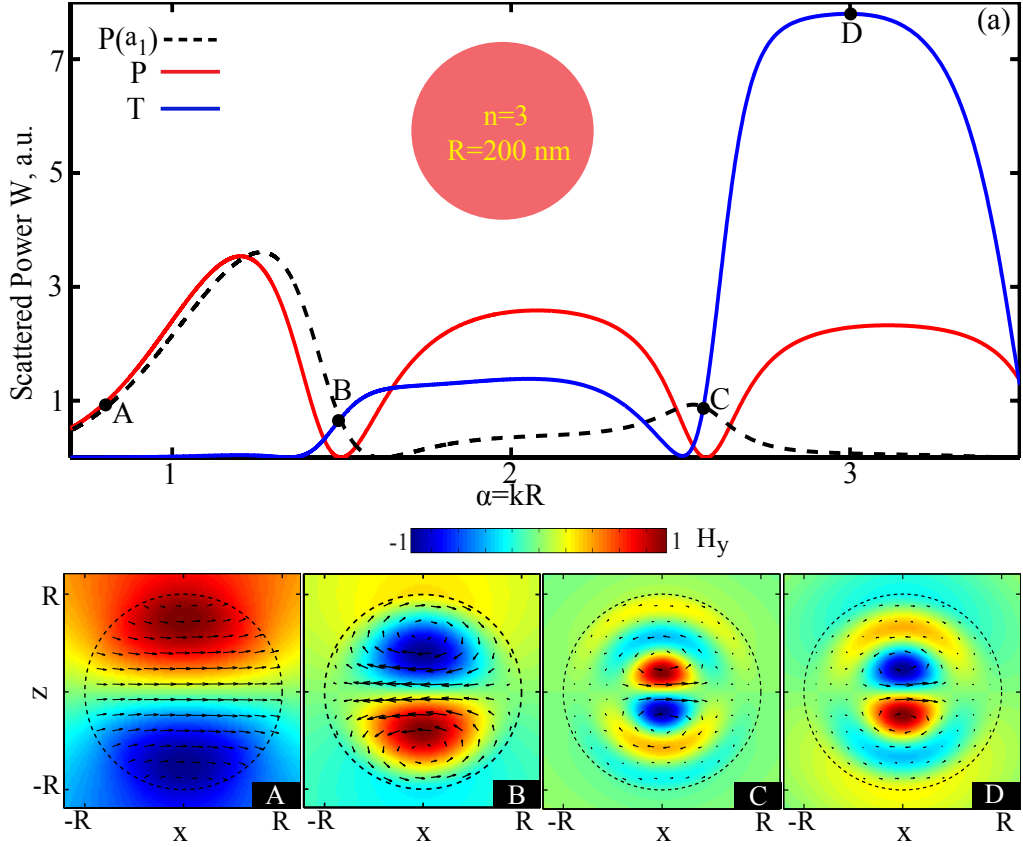}}\caption{\small (a) Scattered power spectra for a dielectric ($n=3$) sphere (inset) of $R=200$~nm. The contributions from current-integrated [based on  Eq.(\ref{ED}) and  Eq.(\ref{TD})] electric dipole ($\textbf{P}$, red curves) and TD ($\textbf{T}$, blue curves), and those from the far-field deduced dipole [$\rm{\textbf{P}(a_1)}$, dashed black curves] are shown. Four points of different $\alpha=kR$ are indicated by black dots in (a) (A: $\alpha=0.8$, B: $\alpha=1.5$, C: $\alpha=2.6$, D: $\alpha=3$). The the corresponding near fields on the $x-z$ plane of $y=0$ are shown respectively in (A)-(D) in the bottom row.  The distributions for both the normalized magnetic field along $y$ direction $\textbf{H}_y$ (color-plots) and displacement field $\textbf{D}=\varepsilon\textbf{E}$ on plane (vector-plots, only the field inside the particle is plotted) are shown (the dashed black lines denote the boundaries of the particles), as is also the case for Figs.~\ref{fig3}-\ref{fig4}.}
\label{fig2}
\end{figure*}

As a first step we study the plane wave scattering by a homogenous and isotropic dielectric sphere of $R=200$~nm and $n=3$. The scattered power spectra [in terms of $\textbf{P}(a_1)$, $\textbf{P}$ and $\textbf{T}$] are shown in Fig.~\ref{fig2}(a). Four points A-D are selected and indicated by black dots in  Fig.~\ref{fig2}(a) and the corresponding near-field distributions are shown in Fig.~\ref{fig2}(A-D): the vector-plots correspond to the displacement field $\textbf{D}=\varepsilon\textbf{E}$ on the $x-z$ plane [only the field within the particle is plotted, as is also the case for Fig.~(\ref{fig2})-Fig.~(\ref{fig3})] and the color-plots correspond to the normalized magnetic fields along $y$ direction ($\textbf{H}_y$).

At point A of $\alpha=0.8$, the quasi-static approximation can be applied, and thus the total dipolar field can be approximated as a conventional  electric dipole: $\textbf{P}(a_1) \approx \textbf{P}$. This is due to the fact that in the quasi-static spectral regime,  $\textbf{T}$ and other higher order correcting terms in Eq.~(\ref{dipole_expansion}) can be neglected. To further confirm this, we show the near fields at point A in Fig.~\ref{fig2}(A), where a typical electric dipole field is exhibited.

\begin{figure*}
\centerline{\includegraphics[width=13.5cm]{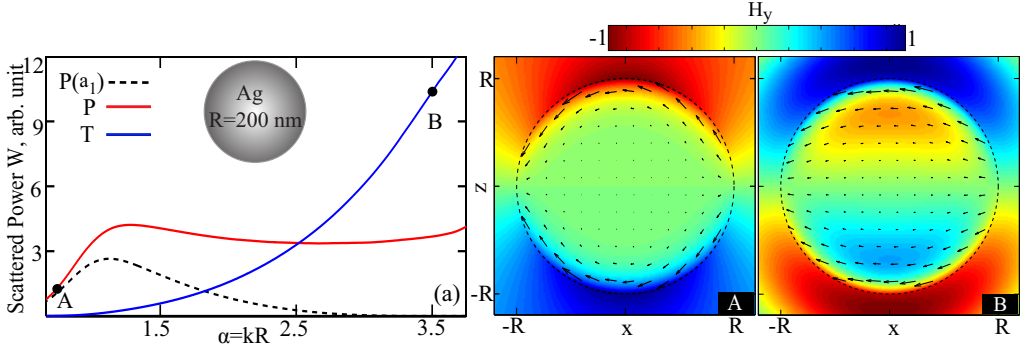}}\caption{\small (a) Scattered power spectra for a silver sphere (inset) of $R=200$~nm. The contributions from $\textbf{P}$, $\textbf{T}$, and $\rm{\textbf{P}(a_1)}$ are shown. Two points of different $\alpha$ are indicated by black dots in (a) (A: $\alpha=0.8$ and B: $\alpha=3.5$). The the corresponding near fields are shown in (A)-(B) on the right.}
\label{fig3}
\end{figure*}

At Points B and C of $\alpha=0.8$ and $2.6$ respectively, which is beyond the quasi-static spectral regime, the quasi-static approximation breaks down and thus the contributions of TD have to be included [see Eq.~(\ref{dipole_expansion})]. At those points, $\textbf{P}$ been totally suppressed and there are only efficient TD excitation: $\textbf{P}(a_1) \approx ik\textbf{T}$.  The corresponding near-field distributions are shown respectively in Fig.~\ref{fig2}(B-C). It is worth mentioning that the field distribution at points B in Fig.~\ref{fig2}(B) corresponds exactly to the filed and currents shown in Fig.~\ref{fig1}(b), which is a typical and fundamental representation of TD [the magnetic field is confined within the particle and there are two field intensity peaks (thus two oppositely circulating current loops on plane) across]. The field distribution at point C shown in Fig.~\ref{fig2}(C), though different from what is shown in Fig.~\ref{fig1}(B), also corresponds to a TD, which however is of a higher mode number with more field lobes~\cite{Lam1992_JOSAB_Explicit,Liu2014_arXiv_Geometric,Liu2015_OE_Ultra} [the magnetic field is also confined within the particle while there are four field intensity peaks across].

At Points D of $\alpha=3$ which is far beyond the quasi-static spectral regime, even the incorporation of TD will not account for all the dipolar scattering, which can be  confirmed in Fig.~\ref{fig2}(a): though at point D the overall dipolar scattering is negligible [$\textbf{P}(a_1)$, dashed black curve],  TD response $\textbf{T}$ is dominant over $\textbf{P}$ and this means that they cannot interfere totally in a destructive way, thus catcalling the scattering of each other~\cite{Liu2014_arxiv,Liu2015_OL_invisible}; As a result, besides $\textbf{P}$  and $\textbf{T}$, there must be higher order correcting terms for $\textbf{P}(a_1)$ as shown in Eq.~(\ref{dipole_expansion}).  This further proves that the multipole expansions based on near-field integration is still incomplete even of the toroidal responses have been considered. Besides conventional  electric, magnetic and toroidal multipoles, there are other terms which will become more and more significant as the  particle dimension gets larger and larger compared to the functioning wavelength~\cite{Dubovik1990_PR,Radescu2002_PRE,Corbaton2015_arXiv_On,Corbaton2015_arXiv_Exact}. The corresponding near-field distribution of point D is shown in Fig.~\ref{fig2}(D).

\begin{figure*}
\centerline{\includegraphics[width=13.5cm]{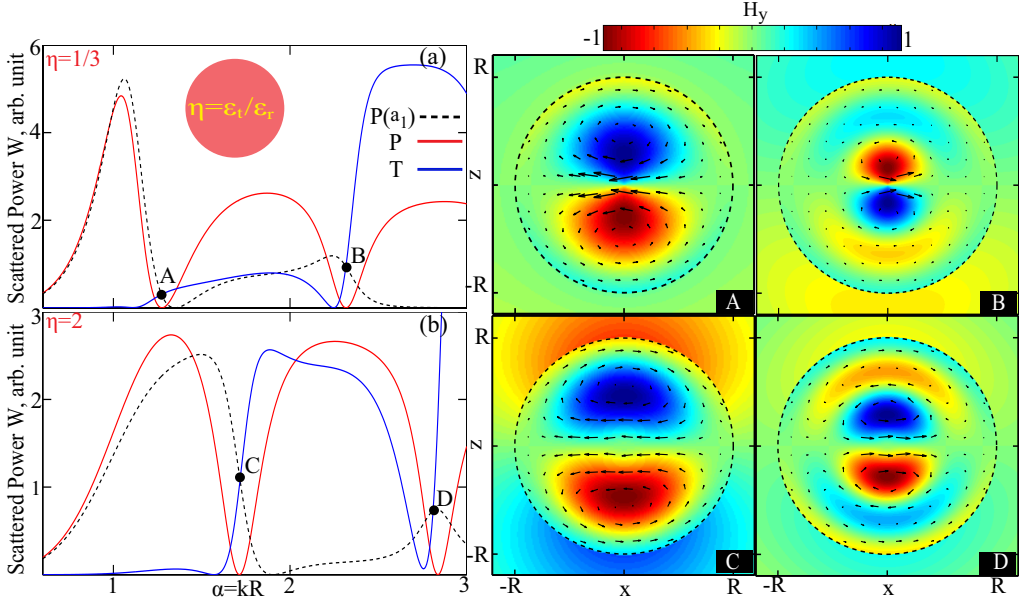}}\caption{\small Scattered power spectra for dielectric spheres (inset) of radius $R=200$~nm with anisotropy parameters  $\eta=\varepsilon_t/\varepsilon_r=1/3$ in (a) and $\eta=2$ in (b). The transverse permittivity is fixed at $\varepsilon_t=9$. The contributions from $\textbf{P}$, $\textbf{T}$, and $\rm{\textbf{P}(a_1)}$ are shown.  Four points are indicated by black dots in (a) and (b) (A: $\alpha=1.27$, B: $\alpha=2.31$, C: $\alpha=1.72$, D: $\alpha=2.8$). The corresponding near fields are shown respectively in (A)-(D) on the right.}
\label{fig4}
\end{figure*}

\subsection{TD excitation within homogenous and isotropic silver particles}
As a next step we study the plane wave scattering by a homogenous and isotropic Ag sphere of $R=200$~nm and the permittivity of silver is adopted from the experimental data in Ref.~\cite{Johnson1972_PRB}. The scattered power spectra [in terms of $\textbf{P}(a_1)$, $\textbf{P}$ and $\textbf{T}$] are shown in Fig.~\ref{fig3}(a). Two points A and B are selected and indicated by black dots in  Fig.~\ref{fig3}(a) and the corresponding near-field distributions are shown in Fig.~\ref{fig3}(A-B). Similar to the dielectric sphere discussed in the subsection above: At point A of $\alpha=0.8$ in the quasi-static regime, $\textbf{P}$ is the only present term [$\textbf{P}(a_1)\approx \textbf{P}$, see Fig.~\ref{fig3}(a)] and the near field exhibits a conventional  electric dipole configuration [see Fig.~\ref{fig3}(A)]; At point B of $\alpha=3.5$ beyond the quasi-static spectral regime, the TD term arises [see Fig.~\ref{fig3}(a)] and the corresponding near field shown in  Fig.~\ref{fig3}(B) is similar to that shown in Fig.~\ref{fig2}(B). The difference is that in Fig.~\ref{fig2}(B) there is only TD excitation while in Fig.~\ref{fig3}(B) there are also $\textbf{P}$ and higher order correcting terms, similar to point D in Fig.~\ref{fig2}(D). In a word, here we show that for homogenous metallic particles, $\textbf{T}$ can be effectively excited beyond the quasi-static spectral regime, which can be stronger over $\textbf{P}$  [Fig.~\ref{fig3}(a)].

\subsection{Tuning the TDs excited within homogenous dielectric particle through radial anisotropy}

At the end, we study the scattering of radially anisotropic dielectric spheres of radius $R=200$~nm.  The anisotropy parameter is $\eta=\varepsilon_t/\varepsilon_r$ and the transverse permittivity is fixed at $\varepsilon_t=9$.  For radially anisotropic particles, both the Mie scattering coefficients and the near-fields can be calculated analytically in a similar manner to that of isotropic particles through modifying the expressions of the orders of related spherical Bessel functions to include the anisotropy parameters~\cite{Qiu2007_PRE_scattering,Qiu2010_IPOR_Light,Ni2013_OE_controlling,Liu2015_OE_Ultra}. As a result, current $\textbf{J}$ [as is shown in Eq.~(\ref{current}) where $\varepsilon(\textbf{r})$  now is a tensor], $\textbf{P}$, $\textbf{T}$, $\rm{\textbf{P}(a_1)}$ and their scattered power for radially anisotropic spherical particles can also be calculated directly.

The scattered power spectra are shown in Fig.~\ref{fig4}(a) of $\eta=1/3$  and in  Fig.~\ref{fig4}(b) of $\eta=2$ with contributions from $\textbf{P}$, $\textbf{T}$, and $\rm{\textbf{P}(a_1)}$. Four points where only TD are excited [$\textbf{P}(a_1)\approx ik\textbf{T}$] with both suppressed $\textbf{P}$ and higher order correcting terms are indicated by black dots (A: $\alpha=1.27$, B: $\alpha=2.31$, C: $\alpha=1.72$, D: $\alpha=2.8$). The  corresponding near fields are shown respectively in Fig.~\ref{fig4}(A)-(D). It is known that through employing radial anisotropy,  $\rm{\textbf{P}(a_1)}$ can be tuned to larger/smaller wavelengths (smaller/larger $\alpha$) through employing smaller/larger $\eta$~\cite{Liu2015_OE_Ultra}. Similarly, TD excited within dielectric spheres can also be tuned by adjusting the radial anisotropy parameters, which is clear according to Fig.~\ref{fig4}(a) and (b).  Compared to the isotropic case of $\eta=1$ shown in Fig.~\ref{fig2}(a), the TD responses [\textit{e.g.} characterized by the positions where only TD is excited, as indicated by points B and C in Fig.~\ref{fig2}(a) and points A-D in Fig.~\ref{fig4}(a-b)] can be tuned to larger wavelengths (smaller $\alpha$) through making $\eta<1$  [see Fig.~\ref{fig4}(a)] and to smaller wavelengths (larger $\alpha$) through rendering $\eta>1$ [Fig.~\ref{fig4}(b)]. For the near-field distributions, both TD with low mode number [see Fig.~\ref{fig4}(A) and (C), which is similar to Fig.~\ref{fig2}(B)] and higher mode number [see Fig.~\ref{fig4}(B) and (D), which is similar to Fig.~\ref{fig2}(C)] can be effectively tuned within radially anisotropic dielectric particles.

\section{Conclusions and Outlook}

To conclude, we revisit the fundamental problem of Mie scattering of spherical particles from the new perspective of TD excitation and tuning.  Besides the demonstration of TDs within dielectric particles, we further demonstrate that TDs can be effectively excited even within homogenous metallic particles as long as it is beyond the quasi-static spectral regime. It is further shown that the TDs excited can be spectrally tuned by adjusting the radial anisotropy parameters, which paves the way for further flexible manipulations of toroidal responses.

It is worth mentioning that here we have confined our study to the electric dipolar responses (transverse magnetic dipolar responses~\cite{Bohren1983_book,Liu2014_arxiv}) and similar studies can certainly conducted for the magnetic dipolar responses (transverse electric dipolar responses~\cite{Bohren1983_book,Liu2014_arxiv}), or both transverse magnetic and transverse electric responses of higher orders.  Such multipole expansions including toroidal responses can also be implemented for other non-spherical structures where the fields and currents, and thus related multipoles can be calculated through numerical methods.  As for the expansion of the dipolar scattering shown in Eq.~(\ref{dipole_expansion}), we have only discussed in detail the correcting terms up to toroidal dipoles, and similarly higher order correcting terms can also be further investigated~\cite{Dubovik1990_PR,Radescu2002_PRE,miroshnichenko2014seeing,Liu2014_arxiv,Liu2015_OL_invisible,Corbaton2015_arXiv_On,Corbaton2015_arXiv_Exact}. For the tuning of TDs excited, Here in this paper we employ only the radial anisotropy, and it is natural to expect that the other kinds of anisotropy, nonlinear and/or gain effects can also be used for further more flexible manipulations of toroidal multipole responses.

We believe that the incorporation of toroidal multipoles into scattering nanoparticles and the explorations into their tunability have not only expanded our understanding of the fundamental Mie scattering physics, but also would be of great significance for both far-field scattering pattern shaping~\cite{Liu2014_CPB,Liu2012_ACSNANO,Liu2014_ultradirectional} and near-field manipulations, which could possibly stimulate many scattering related fundamental researches and applications such as nano-lasing~\cite{Oulton2009_nature,Noginov2009_nature}, sensing~\cite{Kabashin2009_NM}, nanoantennas~\cite{Novotny2011_NP}, and photovoltaic devices~\cite{Atwater2010_NM}.   Moreover, investigations into light-matter interactions involving toroidal multipoles can be possibly conducted on the recently emerging topological~\cite{Lu2014_topological} and/or two-dimensional~\cite{Xia2014_2D} photonic platforms, and probably can lay a solid foundation for the decisive solution of the dynamical Aharonov-Bohm effect controversy concerning quantum information processing~\cite{Afanasiev1995,Marengo2002_Nonradiating}.

 \section*{Acknowledgments}
We thank Jianfa Zhang  and Yuri S. Kivshar for valuable discussions, and acknowledge the financial support from the National Natural Science Foundation of China (Grant numbers: $11404403$ and $61205141$), the Australian Research Council (Grant number: FT110100037) and the Basic Research Scheme of College of Optoelectronic Science and Engineering, National University of Defense Technology.

\end{sloppy}

\begin{thebibliography}{10}
\newcommand{\enquote}[1]{``#1''}

\bibitem{Kaelberer2010_Science}
T.~Kaelberer, V.~Fedotov, N.~Papasimakis, D.~Tsai, and N.~Zheludev,
  \enquote{Toroidal dipolar response in a metamaterial,} Science \textbf{330},
  1510 (2010).

\bibitem{dong2012toroidal}
Z.~Dong, P.~Ni, J.~Zhu, X.~Yin, and X.~Zhang, \enquote{Toroidal dipole response
  in a multifold double-ring metamaterial,} Opt. Express \textbf{20}, 13065
  (2012).

\bibitem{Huang2012_OE}
Y.-W. Huang, W.~T. Chen, P.~C. Wu, V.~Fedotov, V.~Savinov, Y.~Z. Ho, Y.-F.
  Chau, N.~I. Zheludev, and D.~P. Tsai, \enquote{Design of plasmonic toroidal
  metamaterials at optical frequencies,} Opt. Express \textbf{20}, 1760 (2012).

\bibitem{Ogut2012_NL}
B.~Ogut, N.~Talebi, R.~Vogelgesang, W.~Sigle, and P.~A. van Aken,
  \enquote{Toroidal plasmonic eigenmodes in oligomer nanocavities for the
  visible,} Nano Lett. \textbf{12}, 5239 (2012).

\bibitem{Fedotov2013_SR}
V.~A. Fedotov, A.~Rogacheva, V.~Savinov, D.~Tsai, and N.~I. Zheludev,
  \enquote{Resonant transparency and non-trivial non-radiating excitations in
  toroidal metamaterials,} Sci. Rep. \textbf{3}, 2967 (2013).

\bibitem{Fan2013_lowloss}
Y.~Fan, Z.~Wei, H.~Li, H.~Chen, and C.~M. Soukoulis, \enquote{Low-loss and
  high-q planar metamaterial with toroidal moment,} Phys. Rev. B \textbf{87},
  115417 (2013).

\bibitem{savinov2014toroidal}
V.~Savinov, V.~Fedotov, and N.~Zheludev, \enquote{Toroidal dipolar excitation
  and macroscopic electromagnetic properties of metamaterials,} Phys. Rev. B
  \textbf{89}, 205112 (2014).

\bibitem{Xiao2014_AP_core}
Q.~Zhang, J.~J. Xiao, X.~M. Zhang, D.~Han, and L.~Gao,
  \enquote{Core-shell-structured dielectric-metal circular nanodisk antenna:
  Gap plasmon assisted magnetic toroid-like cavity modes,} ACS Photon.
  \textbf{2}, 60 (2014).

\bibitem{Basharin2014_arXiv}
A.~A. Basharin, M.~Kafesaki, E.~N. Economou, C.~M. Soukoulis, V.~A. Fedotov,
  V.~Savinov, and N.~I. Zheludev, \enquote{Dielectric metamaterials with
  toroidal dipolar response,} Phys. Rev. X \textbf{5}, 011036 (2015).

\bibitem{miroshnichenko2014seeing}
A.~E. Miroshnichenko, A.~B. Evlyukhin, Y.~F. Yu, R.~M. Bakker, A.~Chipouline,
  A.~I. Kuznetsov, B.~Lukyanchuk, B.~N. Chichkov, and Y.~S. Kivshar,
  \enquote{Seeing the unseen: observation of an anapole with dielectric
  nanoparticles,} Nat. Commun.  (2015).  arXiv:1412.0299.

\bibitem{Liu2014_arxiv}
W.~Liu, J.~Zhang, and A.~E. Miroshnichenko, \enquote{Toroidal dipole induced
  transparency for core-shell nanoparticles,} Laser Photon.
  Rev.,doi:10.1002/lpor.201500102  (2015). arXiv:1412.4931.

\bibitem{Bakker2015_NL_Magnetic}
R.~M. Bakker, D.~Permyakov, Y.~F. Yu, D.~Markovich, R.~Paniagua-Dom¨ªnguez,
  L.~Gonzaga, A.~Samusev, Y.~Kivshar, B.~Luk¡¯yanchuk, and A.~I. Kuznetsov,
  \enquote{Magnetic and electric hotspots with silicon nanodimers,} Nano Lett.
  \textbf{15}, 2137 (2015).

\bibitem{Liu2015_OL_invisible}
W.~Liu, J.~Zhang, B.~Lei, H.~Hu, and A.~E. Miroshnichenko, \enquote{Invisible
  nanowires with interfering electric and toroidal dipoles,} Opt. Lett.
  \textbf{40}, 2293 (2015).

\bibitem{Dubovik1990_PR}
V.~M. Dubovik and V.~V. Tugushev, \enquote{Toroidmoments in electrodynamics and
  solid-state physics,} Phys. Rep. \textbf{187}, 145 (1990).

\bibitem{Radescu2002_PRE}
E.~E. Radescu and G.~Vaman, \enquote{Exact calculation of the angular momentum
  loss, recoil force, and radiation intensity for an arbitrary source in terms
  of electric, magnetic, and toroid multipoles,} Phys. Rev. E \textbf{65},
  046609 (2002).

\bibitem{jackson1962classical}
J.~D. Jackson, \emph{Classical electrodynamics} (Wiley New York, 1962).

\bibitem{Bohren1983_book}
C.~F. Bohren and D.~R. Huffman, \emph{Absorption and Scattering of Light by
  Small Particles} (Wiley, 1983).

\bibitem{Novotny2012_book}
L.~Novotny and B.~Hecht, \emph{Principles of nano-optics} (Cambridge University
  Press, Cambridge, 2012), 2nd ed.

\bibitem{Liu2014_arXiv_Geometric}
W.~Liu, R.~F. Oulton, and Y.~S. Kivshar, \enquote{Geometric interpretations for
  resonances of plasmonic nanoparticles,} Sci. Rep. \textbf{5}, 12148 (2015).

\bibitem{Kuznetsov2012_SciRep}
A.~I. Kuznetsov, A.~E. Miroshnichenko, Y.~H. Fu, J.~B. Zhang, and B.~S.
  Lukyanchuk, \enquote{Magnetic light,} Sci. Rep. \textbf{2}, 492 (2012).

\bibitem{Evlyukhin2012_NL}
A.~B. Evlyukhin, S.~M. Novikov, U.~Zywietz, R.~L. Eriksen, C.~Reinhardt, S.~I.
  Bozhevolnyi, and B.~N. Chichkov, \enquote{Demonstration of magnetic dipole
  resonances of dielectric nanospheres in the visible region,} Nano Lett.
  \textbf{12}, 3749 (2012).

\bibitem{Liu2014_CPB}
W.~Liu, A.~E. Miroshnichenko, and Y.~S. Kivshar, \enquote{Control of light
  scattering by nanoparticles with optically-induced magnetic responses,} Chin.
  Phys. B \textbf{23}, 047806 (2014).

\bibitem{Doyle1989_optical}
W.~T. Doyle, \enquote{Optical properties of a suspension of metal spheres,}
  Phys. Rev. B \textbf{39}, 9852 (1989).

\bibitem{Wheeler2006_PRB}
M.~S. Wheeler, J.~S. Aitchison, and M.~Mojahedi, \enquote{Coated nonmagnetic
  spheres with a negative index of refraction at infrared frequencies,} Phys.
  Rev. B \textbf{73}, 045105 (2006).

\bibitem{Corbaton2015_arXiv_On}
I.~Fernandez-Corbaton, S.~Nanz, and C.~Rockstuhl, \enquote{On the dynamic
  toroidal multipoles,} arXiv:1507.00755  (2015).

\bibitem{Corbaton2015_arXiv_Exact}
I.~Fernandez-Corbaton, S.~Nanz, R.~Alaee, and C.~Rockstuhl, \enquote{Exact
  dipolar moments of a localized electric current distribution,}
  arXiv:1507.00752  (2015).

\bibitem{Lam1992_JOSAB_Explicit}
C.~C. Lam, P.~T. Leung, and K.~Young, \enquote{Explicit asymptotic formulas for
  the positions, widths, and strengths of resonances in mie scattering,} J.
  Opt. Soc. Am. B \textbf{9}, 1585 (1992).

\bibitem{Liu2015_OE_Ultra}
W.~Liu, \enquote{Ultra-directional super-scattering of homogenous spherical
  particles with radial anisotropy,} Opt. Express \textbf{23}, 14734 (2015).

\bibitem{Johnson1972_PRB}
P.~B. Johnson and R.~W. Christy, \enquote{Optical constants of the noble
  metals,} Phys. Rev. B \textbf{6}, 4370 (1972).

\bibitem{Qiu2007_PRE_scattering}
C.-W. Qiu, L.-W. Li, T.-S. Yeo, and S.~Zouhdi, \enquote{Scattering by
  rotationally symmetric anisotropic spheres: Potential formulation and
  parametric studies,} Phys. Rev. E \textbf{75}, 026609 (2007).

\bibitem{Qiu2010_IPOR_Light}
C.~Qiu, L.~Gao, J.~D. Joannopoulos, and M.~Solja{\v{c}}i{\'c}, \enquote{Light
  scattering from anisotropic particles: propagation, localization, and
  nonlinearity,} Laser Photonics Rev. \textbf{4}, 268 (2010).

\bibitem{Ni2013_OE_controlling}
Y.~X. Ni, L.~Gao, A.~E. Miroshnichenko, and C.~W. Qiu, \enquote{Controlling
  light scattering and polarization by spherical particles with radial
  anisotropy,} Opt. Express \textbf{21}, 8091 (2013).

\bibitem{Liu2012_ACSNANO}
W.~Liu, A.~E. Miroshnichenko, D.~N. Neshev, and Y.~S. Kivshar,
  \enquote{Broadband unidirectional scattering by magneto-electric core-shell
  nanoparticles,} ACS Nano \textbf{6}, 5489 (2012).

\bibitem{Liu2014_ultradirectional}
W.~Liu, J.~Zhang, B.~Lei, H.~Ma, W.~Xie, and H.~Hu, \enquote{Ultra-directional
  forward scattering by individual core-shell nanoparticles,} Opt. Express
  \textbf{22}, 16178 (2014).

\bibitem{Oulton2009_nature}
R.~F. Oulton, V.~J. Sorger, T.~Zentgraf, R.~M. Ma, C.~Gladden, L.~Dai,
  G.~Bartal, and X.~Zhang, \enquote{Plasmon lasers at deep subwavelength
  scale,} Nature \textbf{461}, 629 (2009).

\bibitem{Noginov2009_nature}
M.~A. Noginov, G.~Zhu, A.~M. Belgrave, R.~Bakker, V.~M. Shalaev, E.~E.
  Narimanov, S.~Stout, E.~Herz, T.~Suteewong, and U.~Wiesner,
  \enquote{Demonstration of a spaser-based nanolaser,} Nature \textbf{460},
  1110 (2009).

\bibitem{Kabashin2009_NM}
A.~V. Kabashin, P.~Evans, S.~Pastkovsky, W.~Hendren, G.~A. Wurtz, R.~Atkinson,
  R.~Pollard, V.~A. Podolskiy, and A.~V. Zayats, \enquote{Plasmonic nanorod
  metamaterials for biosensing,} Nat. Mater. \textbf{8}, 867 (2009).

\bibitem{Novotny2011_NP}
L.~Novotny and N.~van Hulst, \enquote{Antennas for light,} Nat. Photon.
  \textbf{5}, 83 (2011).

\bibitem{Atwater2010_NM}
H.~A. Atwater and A.~Polman, \enquote{Plasmonics for improved photovoltaic
  devices,} Nat. Mater. \textbf{9}, 865 (2010).

\bibitem{Lu2014_topological}
L.~Lu, J.~D. Joannopoulos, and M.~Soljacic, \enquote{Topological photonics,}
  Nat. Photon. \textbf{8}, 821 (2014).

\bibitem{Xia2014_2D}
F.~Xia, H.~Wang, D.~Xiao, M.~Dubey, and A.~Ramasubramaniam,
  \enquote{Two-dimensional material nanophotonics,} Nat. Photon. \textbf{8},
  899 (2014).

\bibitem{Afanasiev1995}
G.~N. Afanasiev and Y.~P. Stepanovsky, \enquote{The electromagnetic field of
  elementary time-dependent toroidal sources,} J. Phys. A Math. Gen.
  \textbf{28}, 4565 (1995).

\bibitem{Marengo2002_Nonradiating}
E.~A. Marengo and R.~W. Ziolkowski, \enquote{Nonradiating sources, the
  aharonov-bohm effect, and the question of measurability of electromagnetic
  potentials,} Radio Sci. \textbf{37}, 19 (2002).

\end{thebibliography}
\end{document}